\documentclass[12pt]{article}%
\usepackage{amsmath}
\usepackage{url}
\usepackage{float}
\usepackage{graphicx}%
\usepackage{amsfonts}%
\usepackage{amssymb}

\begin{document}

\title{Lagrangian Densities and Principle of Least Action in\ Nonrelativistic Quantum
Mechanics }
\author{Donald H. Kobe\\Department of Physics\\University of North Texas\\Denton, Texas 76203-1427\\USA}
\maketitle

\begin{abstract}
The Principle of Least Action is used with a simple Lagrangian density,
involving second-order derivatives of the wave function, to obtain the
Schr\"{o}dinger \vspace{0in}equation. A Hamiltonian density obtained from this
simple Lagrangian density shows that Hamilton's equations also give the
Schr\"{o}dinger \vspace{0in}equation. This simple Lagrangian density is
equivalent to a standard Lagrangian density with first-order derivatives. For
a time-independent system the Principle of Least Action reduces to the energy
variational principle. For time-dependent systems the Principle of Least
Action gives time-dependent approximations. Using a Hartree product trial wave
function for a time-dependent many-boson system, we apply the Principle of
Least Action to obtain the Gross-Pitaevskii equation that describes a
Bose-Einstein condensate.

\end{abstract}

\section{Introduction}

The Lagrangian approach using Hamilton's Principle of Least (or \vspace
{0in}Stationary) Action has been a unifying principle in almost all areas of
physics to obtain dynamical equations \cite{lanczos}-\cite{moise}.\emph{\ }The
Hamiltonian approach, on the other hand, depends on a Lagrangian to obtain a
canonical momentum and Hamiltonian. \ In nonrelativistic quantum mechanics the
Hamiltonian approach has dominated, however, because the Schr\"{o}dinger
equation uses the Hamiltonian operator obtained by quantizing the canonical
momentum of the particles. \ The wave function in the Schr\"{o}dinger equation
is a complex function in configuration space. From the wave function a
Lagrangian density can be constructed and used in Hamilton's Principle of
Least Action to give the Schr\"{o}dinger \vspace{0in}equation. This same
procedure is used in relativistic quantum field theory, which is primarily
based on the Lagrangian approach because the Lagrangian density is a scalar
density \cite{lurie, weinberg, ryder, moise}. Our approach is different from
the Feynman path integral formulation of quantum mechanics that uses a
classical \emph{particle} Lagrangian in the path integral to obtain the
propagator \cite{feynman}.

In this paper we discuss a simple Lagrangian density for nonrelativistic
quantum mechanics that gives the Schr\"{o}dinger equation when used in the
Principle of Least Action. This Lagrangian density has been called ``seemingly
artificial'' \cite{nesbet} because it involves second-order spatial
derivatives and is apparently complex, but it is nevertheless widely used
\cite{moccia}-\cite{kramer}. From this Lagrangian density we use the canonical
procedure for fields to obtain a canonical momentum conjugate to the wave
function as a generalized coordinate and then find a Hamiltonian density.
Using these, we show that Hamilton's equations also give the \vspace
{0in}Schr\"{o}dinger equation. We then show that this simple Lagrangian
density is equivalent to a standard Lagrangian density that is real and has
only first derivatives \cite{moise}. \ In addition, we show that for
time-independent systems the Principle of Least Action reduces to the energy
variational principle of nonrelativistic quantum mechanics\vspace{0in}
\cite{basbas}. For time-dependent systems the Principle of Least Action can be
used with time-dependent trial wave functions to obtain approximate equations
and solutions. We apply this method to a system of charged bosons with a
Hartree product time-dependent trial wave function to obtain the \vspace
{0in}\vspace{0in}Gross-Pitaevskii equation \cite{gross}-\cite{kobe}, which is
a nonlinear \vspace{0in}Schr\"{o}dinger equation that describes a \vspace
{0in}Bose-Einstein condensate \cite{cornell}.

In Section 2 we review the Hamiltonian for a nonrelativistic quantum system of
many charged particles in an external electromagnetic field. In Section 3 we
give a simple Lagrangian density for the system and show that Hamilton's
Principle of Least Action gives the Schr\"{o}dinger equation. Using this
simple Lagrangian density, we apply the canonical formalism in Section 4 to
obtain a Hamiltonian density and show that Hamilton's equations also give the
Schr\"{o}dinger equation. In Section 5 we give an elementary derivation of the
equation of continuity in configuration space for probability based on the
invariance of the Lagrangian density under infinitesimal global gauge
transformations. We show in Section 6 that our simple Lagrangian density with
second-order spatial derivatives is real and equivalent to the standard one
with first-order derivatives. For a time-independent system we show in Section
7 that Hamilton's Principle reduces to the energy variational principle. For a
time-dependent system we show in Sec. 8 that a time-dependent trial wave
function may be used in Hamilton's Principle and then apply it to obtain a
time-dependent approximate equation for a many-boson system. The conclusion is
given in Section 9.

\section{Hamiltonian for a many-particle system\emph{\ }}

For a system of many charged particles in an electromagnetic field the
Lagrangian of the classical system can be used to obtain a particle
Hamiltonian by using the canonical procedure \cite{goldstein}. \ The system is
quantized by replacing canonical momenta conjugate to the coordinates by
operators satisfying canonical commutation relations \cite{merzbacher} and the
Hamiltonian then becomes an operator. Because of the sum over all particles,
the Hamiltonian operator emphasizes the particle aspect of quantum theory. We
review the Hamiltonian here to establish the background and notation.

The Hamiltonian operator $\hat{H}$ for a system of $N$-particles with masses
$m_{i}$ and charges $q_{i}$ in an external electromagnetic field is%
\begin{equation}
\hat{H}=\sum_{i=1}^{N}\left\{  \frac{1}{2m_{i}}\left[  \mathbf{\hat{p}}%
_{i}\mathbf{-}q_{i}\mathbf{A(r}_{i},t\mathbf{)}\right]  ^{2}+q_{i}%
A_{0}(\mathbf{r}_{i}\mathbf{,}t)\right\}  +V(\mathbf{r}_{1},...,\mathbf{r}%
_{N}), \label{Ham}%
\end{equation}
where $V$ is a conservative potential energy. The canonical momentum operator
$\mathbf{\hat{p}}_{i}\mathbf{=-}i\hbar\mathbf{\nabla}_{i}=-i\hbar
\partial/\partial\mathbf{r}_{i}$ is conjugate to the coordinate $\mathbf{r}%
_{i}$ for particle $i=1,2,\cdot\cdot\cdot,N.$ Together the coordinates and
canonical momenta satisfy the canonical commutation relations. \ The external
classical electromagnetic field is characterized by a vector potential
$\mathbf{A(\mathbf{r}}_{i}\mathbf{,}t\mathbf{)}$ with components $A_{\alpha
}\mathbf{(\mathbf{r}}_{i}\mathbf{,}t\mathbf{)}$ for $\alpha=1,2,3$ and a
scalar potential $A_{0}(\mathbf{r}_{i},t).$ From these potentials the electric
and magnetic fields can be determined in the usual way.

It is convenient for the rest of the paper to use a more compact notation for
the Hamiltonian (\ref{Ham}). The $\alpha$-component of the displacement
$\mathbf{r}_{i}$\ of particle $i$ has components $x_{i\alpha}$\ for
$\ \alpha=1,2,3.$ The $\alpha$-component of the \emph{mechanical }momentum
operator $\mathbf{\hat{P}}_{i}$ for particle $i$ is
\begin{equation}
\hat{P}_{i\alpha}=\hat{p}_{i\alpha}-q_{i}A_{\alpha}, \label{MV}%
\end{equation}
where the $\alpha$-component of the \emph{canonical} momentum operator is
$\hat{p}_{i\alpha}=$ $-i\hbar\partial/\partial x_{i\alpha}$ $=-i\hbar
\partial_{i\alpha}$. Using this notation for mechanical momentum, we can
rewrite the Hamiltonian operator (\ref{Ham}) as%

\begin{equation}
\hat{H}=\frac{1}{2m_{i}}\hat{P}_{i\alpha}^{2}+q_{i}A_{0}(\mathbf{r}_{i},t)+V,
\label{H}%
\end{equation}
where summation over repeated particle numbers $i$ from $1$ to $N$ and
repeated component indices $\alpha$ from $1$ to $3$ is understood. \ 

\section{Lagrangian density and Hamilton's Principle}

The dynamical equations in almost all areas of physics can be obtained from
Hamilton's Principle of \ Least Action \cite{lanczos, yourgrau}. \ For a
many-particle system the wave function $\Psi$ in nonrelativistic quantum
mechanics is a complex function in a $3N$-dimensional configuration space. In
order to use Hamilton's Principle of Least Action to obtain the \vspace
{0in}Schr\"{o}dinger equation we need a Lagrangian density. \ Such a
Lagrangian density can be constructed from the wave function, its complex
conjugate, and their partial derivatives to any order \cite{brown}. \ Because
it involves the wave function, the Lagrangian density emphasizes the wave
aspect of quantum theory. The wave function of the nonrelativistic quantum
system of $N$\ particles is a complex, time-dependent function $\Psi=$
$\Psi(t)=\Psi(\mathbf{r},t)=\Psi(\mathbf{r}_{1},\mathbf{r}_{2},...,\mathbf{r}%
_{N},t),$ where the $3N$-dimensional vector $\mathbf{r}=(\mathbf{r}%
_{1},\mathbf{r}_{2},...,\mathbf{r}_{N})$ is a vector in configuration space.

We postulate a simple Lagrangian density $\mathcal{L}$\ for this system in
terms of the nonrelativistic wave function $\Psi$ as%
\begin{equation}
\mathcal{L}=\Psi^{\ast}(i\hbar\partial_{t}-\hat{H})\Psi, \label{L}%
\end{equation}
which depends only on $\Psi^{\ast}$, $\Psi$ and its partial derivatives
$\partial_{t}\Psi=\partial\Psi/\partial t,$\ $\partial_{i\alpha}\Psi
=\partial\Psi/\partial x_{i\alpha}$, and $\partial_{i\alpha}^{2}\Psi
=\partial^{2}\Psi/\partial x_{i\alpha}^{2}$. \ Any Lagrangian density that
gives the correct equation of motion is a valid one, so a Lagrangian density
that is complex can still be useful \cite{holland}. The Lagrangian density
need not have only first-order derivatives as is sometime thought, but can
have derivatives of any order \cite{brown}. Nesbet \cite{nesbet} has called
the Lagrangian density (\ref{L}) ``seemingly artificial,'' but it is
nevertheless widely used \cite{moccia}-\cite{kramer}.\ The Lagrangian density
for the Dirac equation is similar in form \cite{lurie}.

To obtain the dynamical equations from a Lagrangian density we use Hamilton's
Principle of Least Action. Hamilton's Principle \cite{moise} states that the
action functional $S[\Psi^{\ast},\Psi]$ for any Lagrangian density
$\mathcal{L}$ is stationary,
\begin{equation}
S[\Psi^{\ast},\Psi]=\int dt\int d^{3N}r\;\mathcal{L}(\Psi,\Psi^{\ast
},...)=\text{Stationary,} \label{Act}%
\end{equation}
where the integration is over all time and all configuration space with an
element of volume \ $d^{3N}r=d^{3}r_{1}\cdot\cdot\cdot d^{3}r_{N}.$ The action
(\ref{Act}) is \vspace{0in}stationary when its variation with respect to
either $\Psi^{\ast}$\ or $\Psi$\ \vspace{0in}\ (or both) is zero. For an
arbitrary Lagrangian density $\mathcal{L}(\Psi,\Psi^{\ast},...)$ variation
with respect to $\Psi^{\ast}$ gives the Euler-Lagrange equation \cite{ryder}.
Boundary conditions at infinity on the variations are chosen such that
$\delta\Psi^{\ast}$ and $\delta\Psi$ vanish sufficiently rapid as \emph{%
$\vert$%
}$t|\rightarrow\infty$\ and $r_{i}\rightarrow\infty$\ for all $i.$

When the specific Lagrangian density $\mathcal{L}$ in Eq. (\ref{L}) is
substituted into Eq.\ (\ref{Act}) and variation is made with respect to
$\Psi^{\ast}$, we obtain%

\begin{equation}
\delta S[\Psi^{\ast},\Psi]=\int dt\int d^{3N}r\;\delta\Psi^{\ast}\left(
i\hbar\partial_{t}-\hat{H}\right)  \Psi=0. \label{VAct}%
\end{equation}
Since the variation $\delta\Psi^{\ast}$ is arbitrary except for vanishing at
the boundaries, Eq. (\ref{VAct}) immediately gives the Schr\"{o}dinger
\vspace{0in}equation%

\begin{equation}
\hat{H}\Psi=i\hbar\frac{\partial\Psi}{\partial t}, \label{SE}%
\end{equation}
where the wave function $\Psi(0)$ must be specified. If the variation of the
action is made with respect to $\Psi,$\ then integration by parts is necessary
and the complex conjugate of the Schr\"{o}dinger \vspace{0in}equation is obtained.

\section{Hamiltonian density and Hamilton's equations}

From the Lagrangian density in Eq. (\ref{L}) we can obtain the Hamiltonian
density by the canonical procedure. \ The Hamiltonian density can be used to
obtain Hamilton's equations. To be consistent with the Principle of Least
Action, Hamilton's equations must also give the Schr\"{o}dinger \vspace
{0in}equation (\ref{SE}), and indeed they do. This section uses functional
derivatives and is more detailed for pedagogical purposes. \ 

The definition of the Hamiltonian from the Lagrangian for classical particles
may be generalized for fields \cite{goldstein}. \ The many-particle wave
function $\Psi(\mathbf{r},t),$ where time $t$ is a parameter and $\mathbf{r}$
is a vector in the $3N$-dimensional configuration space, can be considered as
a generalized coordinate with a conjugate momentum $\Pi(\mathbf{r},t)$ in the
canonical formalism. The\ Hamiltonian density $\mathcal{H}$ is given by a
Legendre transformation \cite{goldstein} to obtain a function of $\Psi(t)$ and
$\Pi(t)$ from the Lagrangian density $\mathcal{L}$ (\ref{L}),
\begin{equation}
\mathcal{H}(\Pi(t),\Psi(t))=\Pi(\mathbf{r},t)\dot{\Psi}(\mathbf{r}%
,t)-\mathcal{L}(\Psi(t),\dot{\Psi}(t)), \label{LH}%
\end{equation}
where $\dot{\Psi}(\mathbf{r},t)\equiv\partial_{t}\Psi(\mathbf{r},t)$.\ The
canonical momentum $\Pi(\mathbf{r},t)$ conjugate to the generalized coordinate
$\Psi(\mathbf{r},t)$ is chosen to eliminate the coefficient of $\delta
\dot{\Psi}$ in the variation of Eq. (\ref{LH}), which gives%
\begin{equation}
\Pi(\mathbf{r},t)\equiv\frac{\partial\mathcal{L}}{\partial\dot{\Psi
}(\mathbf{r},t)}=i\hbar\Psi^{\ast}(\mathbf{r},t). \label{canmom}%
\end{equation}
\qquad

Substituting $\Pi(\mathbf{r},t)=i\hbar\Psi^{\ast}(\mathbf{r},t)$ and
$\mathcal{L}$ in Eq. (\ref{L}) into Eq. (\ref{LH}), we find that the
Hamiltonian density is $\mathcal{H}=\Psi^{\ast}\hat{H}\Psi$, which is\ the
density of the Hamiltonian operator $\hat{H}$ in Eq. (\ref{Ham}). The spatial
integral of the Hamiltonian density is the energy $\mathcal{E}(t)$ of the system%

\begin{equation}
\mathcal{E}(t)\mathbb{=}\int d^{3N}r\text{ }\mathcal{H}=\int d^{3N}r\text{
}\Psi^{\ast}(\mathbf{r},t)\hat{H}\Psi(\mathbf{r},t). \label{Energy}%
\end{equation}
However, when the integral of the Hamiltonian density in Eq. (\ref{Energy}) is
written in terms of the canonical momentum $\Pi(\mathbf{r},t)=i\hbar\Psi
^{\ast}(\mathbf{r},t)$ it can be considered as a Hamiltonian functional
$\mathbb{H[}\Psi(t),\Pi(t)]$ for the system
\begin{equation}
\mathbb{H[}\Psi(t),\Pi(t)]\mathbb{=}\int d^{3N}r\text{ }\mathcal{H}%
(t)=\frac{1}{i\hbar}\int d^{3N}r\text{ }\Pi(\mathbf{r},t)\hat{H}\Psi
(\mathbf{r},t). \label{Hfuncal}%
\end{equation}
From the Hamiltonian functional and the Euler-Lagrange equation Hamilton's
equations \cite{moise, doughty} may be written in a familiar form using
\emph{functional }derivatives.

Hamilton's first equation for fields is \cite{weinberg} \emph{ }
\begin{equation}
\dot{\Psi}(\mathbf{x},t)=\frac{\delta\mathbb{H[}\Psi,\Pi]}{\delta
\Pi(\mathbf{x},t)}, \label{Heq1}%
\end{equation}
where the left-hand side is $\dot{\Psi}=\partial_{t}\Psi=\partial\Psi/\partial
t.$ The right-hand side is the functional derivative of the Hamiltonian
functional $\mathbb{H[}\Psi,\Pi]$ with respect to the canonical momentum
$\Pi(\mathbf{x},t),$where $\mathbf{x=(x}_{1},\mathbf{x}_{2},...,\mathbf{x}%
_{N})$ is also a $3N$-dimensional vector in configuration space. When the
Hamiltonian functional in Eq. (\ref{Hfuncal}) is substituted into it,
Hamilton's first equation gives
\begin{align}
\dot{\Psi}(\mathbf{x},t)  &  =\frac{\delta}{\delta\Pi(\mathbf{x},t)}\int
d^{3N}r\frac{1}{i\hbar}\Pi(\mathbf{r},t)\hat{H}\Psi(\mathbf{r},t),\nonumber\\
\partial_{t}\Psi(\mathbf{x},t)  &  =\frac{1}{i\hbar}\int d^{3N}r\frac{\delta
\Pi(\mathbf{r},t)}{\delta\Pi(\mathbf{x},t)}\hat{H}\Psi(\mathbf{r}%
,t),\nonumber\\
\frac{\partial\Psi(\mathbf{x},t)}{\partial t}  &  =\frac{1}{i\hbar}%
\hat{H}\Psi(\mathbf{x},t), \label{h1}%
\end{align}
which is the \vspace{0in}Schr\"{o}dinger equation (\ref{SE}). \ The functional
derivative of $\Pi(\mathbf{r},t)\mathbb{\ }$[or $\Psi(\mathbf{r}%
,t)$]$\mathbb{\ }$with respect to $\Pi(\mathbf{x},t)$ [or $\Psi(\mathbf{x}%
,t)$] in Eq. (\ref{h1}) is a $3N$-dimensional delta function \cite{ryder},%

\begin{equation}
\frac{\delta\Pi(\mathbf{r},t)}{\delta\Pi(\mathbf{x},t)}=\delta^{(3N)}%
(\mathbf{r-x}). \label{delta}%
\end{equation}
Integrating over this delta function in Eq. (\ref{h1}), we obtain the
Schr\"{o}dinger equation (\ref{SE}). The dimensions of the delta-function on
the right-hand side of Eq. (\ref{delta}) are $(length)^{-3N},$ so the
functional derivative $\frac{\delta\Pi(\mathbf{r},t)}{\delta\Pi(\mathbf{x}%
,t)}$ also has the same dimensions, contrary to appearances. Thus, it is only
apparent that there is a lack of dimensional consistency in Eqs. (\ref{Heq1})
and (\ref{h1}).

Hamilton's second equation is \cite{weinberg}\emph{ }%
\begin{equation}
\dot{\Pi}(\mathbf{x},t)=-\frac{\delta\mathbb{H[}\Psi,\Pi\mathbb{]}}{\delta
\Psi(\mathbf{x},t)}, \label{Heq2}%
\end{equation}
where the left-hand side is $\dot{\Pi}(x,t)=\partial_{t}\Pi(\mathbf{x}%
,t)=i\hbar\partial\Psi^{\ast}(\mathbf{x},t)/\partial t$ from Eq.
(\ref{canmom}). The right-hand side is the functional derivative of the
Hamiltonian functional with respect to the generalized coordinate
$\Psi(\mathbf{x},t).$ Using the Hermiticity of the Hamiltonian, taking the
functional derivative inside the integral to obtain a delta function and
integrating, we obtain Hamilton's second equation,%
\begin{align}
\dot{\Pi}(\mathbf{x},t)  &  =-\frac{\delta}{\delta\Psi(\mathbf{x},t)}\int
d^{3N}r\frac{1}{i\hbar}\Pi(\mathbf{r},t)\hat{H}\Psi(\mathbf{r},t),\nonumber\\
\partial_{t}\Pi(\mathbf{x},t)  &  =-\frac{1}{i\hbar}\int d^{3N}r\hat{H}\Pi
(\mathbf{r},t)\frac{\delta\Psi(\mathbf{r},t)}{\delta\Psi(\mathbf{x}%
,t)},\nonumber\\
i\hbar\frac{\partial\Psi^{\ast}(\mathbf{x},t)}{\partial t}  &  =-\hat{H}\Psi
^{\ast}(\mathbf{x},t), \label{SECC}%
\end{align}
which is the complex conjugate of the \vspace{0in}Schr\"{o}dinger equation
(\ref{SE}). Therefore, Eqs. (\ref{h1}) and (\ref{SECC}) show that Hamilton's
equations obtained from the canonical formalism using the Hamiltonian density
(\ref{LH}) obtained from the Lagrangian density (\ref{L}) is consistent with
Hamilton's Principle of Least Action in giving the Schr\"{o}dinger equation.

\section{Conservation of probability}

Noether's theorem states that the invariance of a Lagrangian density under a
symmetry operation implies a corresponding conservation law \cite{holland,
hill}. \ In particular the invariance of Lagrangian density (\ref{L}) under
infinitesimal global gauge transformations gives the equation of continuity
for probability \cite{moise, sakurai}.\emph{\ }\ Even though this relation is
well known, the proof of it at an elementary level for nonrelativistic quantum
mechanics in configuration space is not readily available. We give a short
derivation of it here to use in the next section.

First consider the variation of the Lagrangian density (\ref{L}) with respect
to $\delta\Psi$. Using ``differentiation by parts'' $[udv=d(uv)-vdu]$ and the
Schr\"{o}dinger \vspace{0in}equation (\ref{SE}), we obtain
\begin{equation}
\delta\mathcal{L}=i\hbar\partial_{t}\left[  \Psi^{\ast}\delta\Psi\right]
+\frac{i\hbar}{2m_{i}}\partial_{i\alpha}\left[  (\hat{P}_{i\alpha}\Psi)^{\ast
}\delta\Psi+\Psi^{\ast}\hat{P}_{i\alpha}\delta\Psi\right]  . \label{dL2}%
\end{equation}
At this point the variation $\delta\Psi$ is arbitrary. When we obtain
$\delta\Psi$ from an infinitesimal global gauge transformation, we find the
equation of continuity for probability.

An infinitesimal global gauge transformation on the wave function $\Psi$ is
\begin{equation}
\Psi^{\prime}=\exp\left\{  i\delta\Gamma/\hbar\right\}  \Psi=\Psi
+(i\delta\Gamma/\hbar)\Psi+\cdot\cdot\cdot, \label{GGT}%
\end{equation}
where $\delta\Gamma=\Sigma_{i}q_{i}\delta\Lambda$ is an infinitesimal
constant. The vector and scalar potentials are of course unchanged by a
constant gauge function $\delta\Lambda$. The new Lagrangian density
$\mathcal{L}^{\prime}=\exp\left\{  -i\delta\Gamma/\hbar\right\}
\mathcal{L}\exp\left\{  i\delta\Gamma/\hbar\right\}  =\mathcal{L}$ is obtained
using the new wave function $\Psi^{\prime}$ in place of $\Psi$ in Eq.
(\ref{L}), so $\delta\mathcal{L}=\mathcal{L}^{\prime}-\mathcal{L}=0.$

The variation in the wave function $\delta\Psi$ under the transformation
(\ref{GGT}) is $\delta\Psi=\Psi^{\prime}-\Psi=(i\delta\Gamma/\hbar)\Psi$
\cite{schweber}. \ Substituting this $\delta\Psi$ into Eq. (\ref{dL2})
and\ using $\delta\mathcal{L}=0,$ we obtain a generalized equation of
continuity%
\begin{equation}
\partial_{t}\left[  \Psi^{\ast}\Psi\right]  +\frac{1}{2m_{i}}\partial
_{i\alpha}\left[  \Psi(\hat{P}_{i\alpha}\Psi)^{\ast}+\Psi^{\ast}(\hat
{P}_{i\alpha}\Psi)\right]  =0. \label{contpsi}%
\end{equation}
This equation can be rewritten as the equation of local probability
conservation in configuration space%

\begin{equation}
\partial_{t}\rho(\mathbf{r},t)+\partial_{i\alpha}J_{i\alpha}(\mathbf{r},t)=0.
\label{cont}%
\end{equation}
The probability density $\rho(\mathbf{r},t)$ and probability current density
$J_{i\alpha}(\mathbf{r},t)$ in the $3N$-dimensional configuration space are
defined respectively as
\begin{equation}
\rho(\mathbf{r},t)=\Psi^{\ast}(\mathbf{r},t)\Psi(\mathbf{r},t),\text{
\ \ \ }J_{i\alpha}(\mathbf{r},t)=\text{Re }\Psi^{\ast}(\mathbf{r},t)\hat
{v}_{i\alpha}\Psi(\mathbf{r},t), \label{rhoJ}%
\end{equation}
where the velocity operator $\mathbf{\hat{v}}_{i}$ is the mechanical momentum
$\mathbf{\hat{P}}_{i}$ divided by the mass $m_{i}$ for particle $i$,
\emph{i.e.,}\text{\textbf{ }}$\mathbf{\hat{v}}_{i}=\mathbf{\hat{P}}_{i}/m_{i}$.

\section{Standard Lagrangian density}

The ``standard'' Lagrangian density $\mathcal{L}_{1}$ that gives the
Schr\"{o}dinger equation is real and has only first-order derivatives
\cite{moise, holland},%

\begin{align}
\mathcal{L}_{1}  &  =\frac{1}{2}\left[  \Psi^{\ast}(i\hbar\partial_{t}%
\Psi)+\Psi(i\hbar\partial_{t}\Psi)^{\ast}\right] \nonumber\\
&  -\frac{1}{2m_{i}}(\hat{P}_{i\alpha}\Psi)^{\ast}(\hat{P}_{i\alpha}\Psi
)-\Psi^{\ast}\left[  V+q_{i}A_{0}(\mathbf{r}_{i},t)\right]  \Psi, \label{L1}%
\end{align}
where the subscript on $\mathcal{L}_{1}$ emphasizes that it has only
first-order partial derivatives. The Lagrangian density $\mathcal{L}_{1}$ in
Eq. (\ref{L1}) can be rewritten using differentiation by parts as%

\begin{equation}
\mathcal{L}_{1}=\mathcal{L}-i\frac{\hbar}{2}\left\{  \partial_{t}(\Psi^{\ast
}\Psi)+\frac{1}{m_{i}}\partial_{i\alpha}\left(  \Psi^{\ast}\hat{P}_{i\alpha
}\Psi\right)  \right\}  . \label{L21}%
\end{equation}
Even though the Lagrangian density $\mathcal{L}$ in Eq. (\ref{L}) is
apparently complex, its imaginary part is zero, which can be shown by taking
the imaginary part of Eq. (\ref{L21}), \
\begin{equation}
\text{Im }\mathcal{L}=\frac{\hbar}{2}\left\{  \partial_{t}(\Psi^{\ast}%
\Psi)+\frac{1}{m_{i}}\partial_{i\alpha}\text{Re }\left(  \Psi^{\ast}%
P_{i\alpha}\Psi\right)  \right\}  =0, \label{Im}%
\end{equation}
from the equation of continuity (\ref{contpsi}). The real part of
$\mathcal{L}$ in Eq. (\ref{L21}) is
\begin{equation}
\text{Re }\mathcal{L}=\mathcal{L=L}_{\text{\emph{1}}}-\frac{\hbar}{2m_{i}%
}\partial_{i\alpha}\text{Im }(\Psi^{\ast}\hat{P}_{i\alpha}\Psi), \label{Re}%
\end{equation}
\ \ so the Lagrangian densities $\mathcal{L}$ and $\mathcal{L}_{\text{\emph{1}%
}}$ are not equal. Nevertheless, they are \emph{equivalent} to each other in
the sense that they have the same action \cite{holland}. Their difference is a
partial derivative with respect to the spatial coordinates which vanishes
because of boundary conditions when integrated to obtain the action
(\ref{Act}). The same dynamical equation, \emph{viz.}, the Schr\"{o}dinger
\vspace{0in}equation, is therefore obtained from both Lagrangian densities.

Sil \cite{sil} was the first to use the real part of $\mathcal{L}$ as a
Lagrangian density,
\begin{equation}
\mathcal{L}_{\text{\emph{Sil}}}=\text{Re }\Psi^{\ast}\left(  i\hbar
\partial_{t}-\hat{H}\right)  \Psi=\text{Re }\mathcal{L}, \label{LSil}%
\end{equation}
which was later used by Mittleman \cite{mittleman}. It was subsequently used
by McCarroll, \emph{et al}. \cite{mccarroll}, who pointed out that Re
$\mathcal{L=L}$.

\section{Energy Variational Principle}

If the quantum mechanical system is time independent then Hamilton's Principle
reduces to the energy variational principle of time-independent quantum
mechanics \cite{basbas}. The time-independent Schr\"{o}dinger \vspace
{0in}equation can be obtained by separation of the time variable in the
time-dependent Schr\"{o}dinger equation. \ However, the time-independent
Schr\"{o}dinger \vspace{0in}equation can also be obtained from Hamilton's
principle by using it with a trial wave function.

The Principle of Least Action can be used with a trial wave function of the form%

\begin{equation}
\Psi(\mathbf{r}_{1},\mathbf{r}_{2},...\mathbf{r}_{N},t)=\Psi(\mathbf{r}%
_{1},\mathbf{r}_{2},...\mathbf{r}_{N})\exp\left(  -iEt/\hbar-\epsilon
|t|\right)  , \label{tiwf}%
\end{equation}
where $\Psi$ on the right-hand side is time independent, $E$ is the energy and
$\epsilon>0$ is a small parameter that can be taken to be zero at the end of
the calculation. Using this wave function in Eq. (\ref{Act}) for the action
and doing the time integration, we obtain \
\begin{equation}
S[\Psi^{\ast},\Psi]=\frac{1}{\epsilon}\int d^{3N}r\text{ }\Psi^{\ast}\left(
E-\hat{H}\right)  \Psi=\text{Stationary.} \label{Sss}%
\end{equation}
If this equation is multiplied by $-\epsilon<0$ and the constant $E$ is added,
we obtain the energy variational principle for a stationary state%
\begin{equation}
\int d^{3N}r\text{ }\Psi^{\ast}\hat{H}\Psi-E(\int d^{3N}r\text{ }\Psi^{\ast
}\Psi-1)=\text{Stationary,} \label{EV}%
\end{equation}
which is independent of $\epsilon.$ The energy $E$ is a Lagrangian multiplier
that ensures the normalization of the wave function.

If the variation of Eq. (\ref{EV}) is made with respect to the function
$\Psi^{\ast},$ we obtain the time-independent Schr\"{o}dinger \vspace
{0in}equation%
\begin{equation}
\hat{H}\Psi=E\Psi. \label{tiSE}%
\end{equation}
The variation with respect to $\Psi$ gives the complex conjugate of the
time-independent Schr\"{o}dinger \vspace{0in}equation (\ref{tiSE}) after
integration by parts.

Hamilton's Principle of Least action is however more important in obtaining
approximate solutions for the energy and wave function. If the wave
function\ $\Psi$ in Eq. (\ref{EV}) is replaced by a time-independent
normalized trial wave function $\Phi$ with parameters or unknown functions and
an energy $\left\langle \Phi|\hat{H}\Phi\right\rangle =$ $E^{\prime}.$ Then
Eq. (\ref{EV}) can be varied with respect to these parameters or functions to
obtain the lowest energy $E^{\prime}$ with a wave function of the form $\Phi$.
From the standard Rayleigh-Ritz method it is easy to show that the energy
$\left\langle \Phi|\hat{H}\Phi\right\rangle =$ $E^{\prime}$ is an upper bound
to the true ground state energy $E_{0},$ \emph{i.e.,} $E^{\prime}>E_{0}$
\cite{merzbacher}.

\section{Time-dependent approximation for a many-particle system}

Approximate solutions or equations for the time-dependent Schr\"{o}dinger
\vspace{0in}equation (\ref{SE}) can be obtained from the Principle of Least
Action (\ref{Act}) by using the Lagrangian density (\ref{L}) and replacing the
exact wave function $\Psi(t)$ in by a time-dependent trial wave function
$\Phi(t)$,
\begin{equation}
S[\Phi^{\ast},\Phi]=\int dt\int d^{3N}r\Phi^{\ast}(t)\left(  i\hbar
\partial_{t}-\hat{H}\right)  \Phi(t)=\text{Stationary,} \label{plat}%
\end{equation}
and varying with respect to parameters or functions in $\Phi(t).$ Equation
(\ref{plat}) can be used to obtain a time-dependent approximation for any
quantum system regardless of the number of particles.

For $N$-identical particles with $q_{i}=q$ and $m_{i}=m,$ the potential energy
$V$ in the Hamiltonian (\ref{Ham}) is taken to be the sum of one- and two-body
potentials
\begin{equation}
V(\mathbf{r}_{1},\mathbf{r}_{2},...,\mathbf{r}_{N})=\sum_{i=1}^{N}%
V^{(1)}(\mathbf{r}_{i})+\frac{1}{2}\sum_{i\neq j=1}^{N}V^{(2)}(\mathbf{r}%
_{i},\mathbf{r}_{j}), \label{V}%
\end{equation}
respectively, where\thinspace$\mathbf{\ }$the two-body potential is symmetric
$V^{(2)}(\mathbf{r}_{i},\mathbf{r}_{j})=V^{(2)}(\mathbf{r}_{j},\mathbf{r}%
_{i}).$ This form of the Principle of Least Action is applicable to any system
of many identical particles, either bosons or fermions.

We illustrate Hamilton's Principle with an approximate wave function for a
system of many bosons to obtain the \vspace{0in}Gross-Pitaevskii equation
\cite{gross}-\cite{kobe}, which is a nonlinear \vspace{0in}Schr\"{o}dinger
equation commonly used to treat Bose-Einstein condensates \cite{cornell}. \ We
choose a time-dependent trial wave function $\Phi(t)$ for a system of
$N$-identical bosons as a Hartree product of identical, normalized
single-particle functions $\varphi(\mathbf{r},t\mathbf{),}$
\begin{equation}
\Phi(\mathbf{r}_{1},\mathbf{r}_{2},...,\mathbf{r}_{N},t)=\varphi
(\mathbf{r}_{1},t\mathbf{)}\varphi(\mathbf{r}_{2},t\mathbf{)\cdot\cdot\cdot
}\varphi(\mathbf{r}_{N},t\mathbf{),} \label{Phi}%
\end{equation}
which satisfies Bose-Einstein statistics because it is symmetric under
interchange of particles. Substituting Eq. (\ref{Phi}) into the variational
principle (\ref{plat}) and integrating over the coordinates, we obtain the
action functional in terms of single boson wave functions $\varphi$%

\begin{align}
&  S\left[  \varphi^{\ast},\varphi\right] \nonumber\\
&  =N\int dt\int d^{3}r\varphi^{\ast}(\mathbf{r},t)\left\{  i\hbar\partial
_{t}\varphi(\mathbf{r},t)-\frac{1}{2m}\left[  -i\hbar\mathbf{\nabla
-}q\,\mathbf{A}\right]  ^{2}\varphi(\mathbf{r},t)-q\,A_{0}\varphi
(\mathbf{r},t)\right. \nonumber\\
&  \left.  -V^{(1)}(\mathbf{r})\varphi(\mathbf{r},t)-\frac{1}{2}(N-1)\int
d^{3}r^{\prime}V^{(2)}(\mathbf{r,r}^{\prime})\varphi^{\ast}(\mathbf{r}%
^{\prime},t)\varphi(\mathbf{r}^{\prime},t\mathbf{)}\varphi(\mathbf{r}%
,t)\right\} \nonumber\\
&  =\text{\vspace{0in}Stationary,} \label{ActH}%
\end{align}
where the integration is over all time and space.

Variation of this expression with respect to $\varphi^{\ast}$ gives a
time-dependent Hartree approximation known as the \vspace{0in}Gross-Pitaevskii
equation \cite{gross}-\cite{kobe},%
\begin{align}
&  i\hbar\frac{\partial}{\partial t}\varphi(\mathbf{r},t)=\frac{1}{2m}\left[
-i\hbar\mathbf{\nabla-}q\,\mathbf{A(r},t\mathbf{)}\right]  ^{2}\varphi
(\mathbf{r},t)+[q\,A_{0}(\mathbf{r,}t)+V^{(1)}(\mathbf{r})]\varphi
(\mathbf{r},t)\nonumber\\
&  +(N-1)\int d^{3}r^{\prime}V^{(2)}(\mathbf{r,r}^{\prime})\varphi^{\ast
}(\mathbf{r}^{\prime},t)\varphi(\mathbf{r}^{\prime},t\mathbf{)}\varphi
(\mathbf{r},t), \label{GP}%
\end{align}
for a charged many-boson system in an electromagnetic field \cite{reinisch}.
This equation is a nonlinear Schr\"{o}dinger equation for the single boson
wave function $\varphi$, where the nonlinear term is the average potential on
one particle due to all the other particles in the system. The variation may
also be made with respect to $\varphi(\mathbf{r},t)$ to obtain the complex
conjugate of Eq. (\ref{GP}).

\section{Conclusion}

Hamilton's Principle of Least Action is a fundamental principle of physics
that is used to obtain dynamical equations for both nonrelativistic and
relativistic particles and fields \cite{lanczos, yourgrau}. This principle is
applied to nonrelativistic quantum mechanics by considering the wave function
in configuration space as a generalized coordinate and constructing a
Lagrangian density such that Hamilton's Principle gives the \vspace
{0in}Schr\"{o}dinger equation. This approach gives a unified treatment for
both nonrelativistic quantum theory and relativistic quantum field theory
\cite{lurie}, as well as showing their unity with other branches of physics.
The Lagrangian approach is in a sense more fundamental than the Hamiltonian
approach that has dominated the treatment of \vspace{0in}%
nonrelativistic\ \vspace{0in}quantum mechanics because the Hamiltonian must be
derived from a Lagrangian. In relativistic theory the Lagrangian density is
also a scalar density, whereas the Hamiltonian density is not. The two
approaches are, however, complementary in the sense that the Lagrangian
density using the wave function emphasizes the wave aspect of nature, whereas
the Hamiltonian operator with a sum over particles emphasizes its particle aspect.

We use a simple Lagrangian density for the Schr\"{o}dinger equation that is
apparently complex and has second-order spatial derivatives. This Lagrangian
density is actually real because of the equation of continuity for
probability. It is equivalent to the standard Lagrangian density that is real
and has only first-order derivatives \cite{holland}. Two Lagrangian densities
are equivalent if the difference between the two is a sum of spatial or
temporal partial derivatives so they give the same action.

For time-independent quantum systems, the Principle of Least Action reduces to
the energy variational principle of nonrelativistic quantum mechanics. By
minimizing the energy calculated from a trial wave function, it gives an
approximate wave function and an energy that is an upper bound to the exact
ground state energy.

For time-dependent quantum systems, the Principle of Least Action with a
time-dependent trial wave function gives an approximate equation for the trial
wave function. We apply this approach to a many boson system by using a
Hartree product wave function to obtain the time-dependent \vspace
{0in}Gross-Pitaevskii equation \cite{gross}-\cite{kobe}, which is a nonlinear
Schr\"{o}dinger equation that describes a Bose-Einstein condensate
\cite{cornell}. For fermion systems a Slater determinant with time-dependent
single-particle trial wave functions may be used to obtain the time-dependent
Hartree-Fock equations \cite{nesbet}. \ 

The Lagrangian approach to nonrelativistic quantum mechanics described here
would be suitable to supplement a course in quantum mechanics or field theory.
For quantum mechanics it would show how to obtain time-dependent
approximations. For field theory it would be a simple familiar example. For
both its use would show the unity that Hamilton's Principle of Least Action
provides for\ physics \cite{lanczos}-\cite{moise}.

\section{Acknowledgements}

This paper is dedicated to the memory of my father Kenneth A. Kobe
(1905-1958), Professor of Chemical Engineering at the University of Texas,
Austin, who encouraged me to pursue my interest in physics.

I would like to thank Professor Wolfgang P. Schleich and Dr. Reinhold Waiser
for discussions and Professor Marlan O. Scully for his encouragement. This
work was partially supported by a grant from ONR N00014-03-1-0639/TAMU TEES 53494.

\end{document}